\documentclass[superscriptaddress,floatfix,endfloats]{revtex4}
\usepackage{bm}
\usepackage{graphics}

\newcommand{\avg}[1]{\left \langle #1 \right \rangle}
\newcommand{\myfigure}[1]{\resizebox{3in}{!}{\includegraphics{#1}}}
\newcommand{\vol}[1]{\textbf{#1}}

\newcommand{\Pzeta}{\ensuremath{\Delta^2_{\zeta}}}
\newcommand{\vc}[1]{\ensuremath{\mathbf{#1}}}
\newcommand{\like}{\ensuremath{{\mathcal{L}}}}
\newcommand{\mat}[1]{\ensuremath{\mathbf{#1}}}

\begin{document}
\title{Constraining cutoff physics in the cosmic microwave background}
\author{Takemi Okamoto}
\email{tokamoto@oddjob.uchicago.edu}
\affiliation{The Department of Physics and the Center for Cosmological Physics,
 University of Chicago, Chicago, IL 60637}
\author{Eugene A. Lim}
\email{elim@jasmine.uchicago.edu}
\affiliation{The Department of Astronomy and Astrophysics and the Center for
Cosmological Physics, University of Chicago, Chicago, IL 60637}

\begin{abstract}
We investigate the ability to constrain oscillatory features in the primordial
power spectrum using current and future cosmic microwave background 
observations.  In particular, we study the observability of an oscillation
arising from imprints of physics at the cutoff energy scale.  We perform a 
likelihood analysis on the WMAP data set, and find that the current data set
constrains the amplitude of the oscillations to be less than $0.77$ at 
2$\sigma$, consistent with a power spectrum without oscillations.  
In addition, we investigate the fundamental limitations in the measurement
of oscillation parameters by studying the constraints from a 
cosmic variance limited experiment.
We find that such an experiment is capable of constraining the amplitude of 
such oscillations to be below $0.005$, implying that reasonable models 
with cutoff energy scales $\Lambda>200 H_{\text{infl}}$ are unobservable 
through the microwave background.
\end{abstract}
\maketitle 

\section{Introduction}
\label{Sect:Introduction}

Recent observations of cosmic microwave background (CMB) anisotropies
by the Wilkinson Microwave Anisotropy Probe (WMAP)~\cite{Bennett03,Spergel03a} 
have raised the possibility of
determining the initial conditions for structure formation with high
precision.  The most recent data are consistent with a power-law
primordial power spectrum, as predicted by most variants of the
inflationary paradigm~\cite{Peiris03,Guth81,BST83,Starobinsky82}.
At the same time, marginal
detections of a running of the power-law index, as well as glitches in
the temperature power spectrum across several multipoles, suggest that
the primordial power spectrum may deviate from the power-law form, and
the presence of such features may be tested with future experiments
such as the Planck satellite~\cite{Planck} and CMBPol~\cite{CMBPol}.

Previous works have taken model independent approaches to the measurement
of features in the power spectrum.  These parametrize the primordial
power spectrum in band powers or 
wavelets~\cite{MukherjeeWang03,WangMathews02,Wang99,BridleLewis03}, 
or choose a set of window functions to regularize the initial power
spectrum~\cite{Miller02,Tegmark02}.  An alternate, observation-oriented 
approach involves constructing the principal components of the primordial power
spectrum, which are guaranteed to be the modes that best constrain
deviations from the power-law form~\cite{HuOkamoto03}.  Such general methods 
are ideal for situations where the deviations from the power-law form are 
unknown.  However, if a well-motivated form for the deviations exists, 
the constraints can be improved by incorporating the specific form in the 
analysis.

A form of modification that is favored by several mechanisms is that 
of an oscillation on top of the usual power-law form.  Although such a 
modification can arise from interactions of the inflaton with other 
fields~\cite{Burgess02}, we specifically consider the modifications that arise 
from an ambiguity in the vacuum
choice at some high-energy cutoff scale.  The presence of a high-energy
cutoff in the description of fluctuations during inflation naturally leads
to an oscilatory modification
~\cite{Brandenberger01A,Brandenberger01B,Danielsson02A,Easther02,ArmendarizLim03}.  Current data sets allow a very loose limit on a
restricted, two-parameter form of the oscillation~\cite{Elgaroy03}, and an 
oscillation may be favored~\cite{Martin03}, although no constraints have been
derived on the parameters that define the oscillations.  In addition, there is 
some disagreement over whether such effects might be observable by future 
experiments, and parameter forecasts are absent from current 
literature~\cite{Bergstrom02,Elgaroy03}.  Furthermore, the previous studies 
do not account for the full, four-parameter phenomenology that parametrizes 
our ignorance of the physics at the cutoff scale.

In this work, we quantify the limit on such modifications by using 
current CMB data to constrain the parameters that define the oscillations.  
Furthermore, we provide robust forecasts for the oscillation parameters by
considering the constraints from a cosmic variance limited experiment.
We briefly review
the theoretical framework in Sec.~\ref{Sect:TheoreticalFramework}, and
describe the statistical method in Sec.~\ref{Sect:Analysis}.  We present the
bounds from current CMB data in Sec.~\ref{Sect:WMAPBounds}, and the 
limitations due to cosmic variance in Sec.~\ref{Sect:TheoryLimit}.  
We conclude in Sec.~\ref{Sect:Discussion}.

\section{Theoretical Framework}
\label{Sect:TheoreticalFramework}

In the inflationary scenario, inhomogeneities in the matter density
arise due to quantum fluctuations of the inflaton, the scalar field
that induces inflation.  Computations in such a scenario assume that
the quantum field theoretic description of the inflaton remains valid to
arbitrarily small length scales; however, we expect quantum field theory to
break down at some length scale, so that a complete description of the
quantum fluctuations would have to incorporate the unknown physics at the 
cutoff scale.

In the absence of information concerning the nature of the high-energy
cutoff, we assume that the effects of any physics beyond the cutoff is
equivalent to ambiguities in the choice of vacuum at the cutoff
scale.  We describe a model that parametrizes our ignorance about
the cutoff physics, and briefly review the relation between the
primordial power spectrum from inflation and the cosmic microwave
background.

\subsection{Deviations from scale invariance}
\label{Subsect:PkTheory}

In this section we briefly review the theory of
perturbation generation in an inflationary spacetime, and describe a generic
model for deviations from a scale-invariant power spectrum.
A complete treatment can be found in Ref.~\cite{ArmendarizLim03}.  We follow 
the 
notation in Ref.~\cite{ArmendarizLim03} except for a change in metric 
signature.

We begin by considering a theory of inflation with a minimally coupled
scalar field with a potential $V(\phi)$
\begin{equation}
S=\int d^4x\sqrt{-g}\left[\frac{m_p^2}{2}R-\frac{1}{2}(\partial \phi)^2
-V(\phi)\right], \label{Eqn:GravityAction}
\end{equation}
where $m_p^2\equiv (8\pi G)^{-1}$ is the reduced Planck mass.
Since the universe is spatially homogenous and isotropic we will use
the Robertson-Walker metric:
\begin{equation}
ds^2=a^2(\eta)(-d\eta^2+\delta_{ij}dx^i dx^j). 
\label{Eqn:RobertsonWalkerMetric}
\end{equation}
Note that the conformal time $\eta$ is taken to run from $-\infty$ to $0$. 
We assume a spatially flat metric for simplicity.

If, for a certain range of $\phi$, the potential $V(\phi)$ is nearly
constant and thus the field is rolling slowly, then the scalar field
behaves like a cosmological constant and the universe inflates
during this phase. In the slow-roll limit, the potential 
can be characterized by the parameters~\cite{LythLiddleBook} 
\begin{eqnarray}
\epsilon&\equiv&\frac{3\phi'^2/2}{a^2 V +\phi'^2/2}=
\frac{m_p^2}{2}\left (\frac{\partial_\phi V}{V}\right)^2
, \label{Eqn:slowrollepsilon} \\
\delta &\equiv&\frac{{\cal{H}}\phi'-\phi''}{{\cal{H}}\phi^2} = 
m_p^2\frac{\partial^2_\phi V}{V} 
- \frac{m_p^2}{2}\left (\frac{\partial_\phi V}{V}\right)^2
, \label{Eqn:slowrollbdelta}
\end{eqnarray}
with primes denoting derivatives with respect to conformal time.  The 
conditions $\epsilon, \delta \ll 1$ imply that the universe is inflating,
with $\epsilon=\delta=0$ corresponding to pure de Sitter expansion.
In general, $\epsilon$ and $\delta$ can be functions of time, but
in the slow-roll approximation we can take them to be
constants.

Perturbing the metric
\begin{equation}
g_{\mu\nu}dx^{\mu}dx^{\nu}\rightarrow (g_{\mu\nu}+\delta g_{\mu\nu})
dx^{\mu}dx^{\nu}\equiv a^2(\eta)\left[-(1+2\Phi)d\eta^2+(1-2\Phi)
\delta_{ij}dx^i dx^j\right] \label{Eqn:PerturbedMetric}
\end{equation}
and the scalar field
\begin{equation}
\phi\rightarrow\phi+\delta \phi, \label{Eqn:PerturbedScalarField}
\end{equation}
we compute the action for the perturbations to be~\cite{Mukhanov91}
\begin{equation}
S_{\text{pert.}}=\frac{1}{2}\int d^4x \left[v'^2-(\partial v)^2+\frac{z''}{z}v^2\right]. 
\label{Eqn:PerturbationAction}
\end{equation}
Here, $z\equiv \frac{a {\phi}'}{(a'/a)}$, and $\partial$ is the ordinary
partial derivative over the three spatial indices. The dynamical degree of 
freedom of the perturbations, 
$v\equiv a\left(\delta \phi+(\frac{{\phi}'}{a'/a})\Phi\right)$, is a linear
combination of perturbations in the metric and the scalar field.

Varying the action (\ref{Eqn:PerturbationAction}), we obtain the equation of 
motion 
\begin{equation}
v''+\left(-\partial^2-\frac{z''}{z}\right)v=0 
\label{Eqn:EquationOfMotionPerturbation}
\end{equation}
for the perturbation.  We quantize the field $v$ 
by promoting it to an operator $\hat{v}$ which can be expanded into Fourier
modes and their associated raising and lowering 
operators~\cite{Mukhanov91, BirrellDavies}
\begin{equation}
\hat{v}=\frac{1}{\sqrt{2}}\int \frac{d^3k}{(2\pi)^{3/2}}\left(v_k(\eta)
e^{-i\vc{k}\cdot\vc{x}}\hat{a}_{\vc{k}}+
v_k^{*}(\eta)e^{i\vc{k}\cdot\vc{x}}
\hat{a}_{\vc{k}}^{\dagger}\right). \label{Eqn:ModeExpansion}
\end{equation}
The raising and lowering operators obey the commutation relation 
$[\hat{a}_{\vc{k}},\hat{a}_{\vc{k}'}^{\dagger}]=\delta(\vc{k}-\vc{k}')$ 
if the mode functions satisfy the following normalization condition
\begin{equation}
v_k' v^{*}_k-{v'}_k^{*} v_k=-2i. \label{Eqn:ModeNormalization}
\end{equation}
Substituting Eq. (\ref{Eqn:ModeExpansion}) into the equation of motion
(\ref{Eqn:EquationOfMotionPerturbation}), we obtain the the equation
of motion for the mode function
\begin{equation}v''_k+\left(k^2-\frac{z''}{z}\right)v_k=0. 
\label{Eqn:ModeEquationOfMotion}
\end{equation}

The quantum field theoretic description is incomplete without a specification 
for the
choice of a vacuum state, defined as a state $|0\rangle$ annihilated by 
the lowering operator ($\hat{a}_{\vc{k}}|0\rangle=0$).  
The choice of a vacuum state 
$|0\rangle$ of the field is equivalent to choosing the initial conditions for 
the mode functions $v_k$ at some time $\eta_0$~\cite{BirrellDavies}.  In 
Minkowski spacetime, the vacuum choice corresponding to the initial conditions
\begin{equation}
v_k(\eta_0)=\frac{1}{\sqrt{k}}\ ,\ v'_k(\eta_0)=-i\sqrt{k},
\label{Eqn:MinkowskiVacuum}
\end{equation} 
is the unique choice that is invariant under the symmetries of 
Minkowski spacetime.  In curved spacetimes, however, no unique choice
exists in general.  However, the success of quantum field theory in predicting
results of accelerator experiments implies that length scales much smaller than
the curvature of spacetime, the vacuum is essentially the same as the Minkowski
vacuum.  We therefore choose a parametrized set of vacua that reduce to the 
Minkowski vacuum when the curvature $H^{-1}$ is small compared to the physical 
length of the mode $k$.

Following Ref.~\cite{ArmendarizLim03}, we impose the initial conditions
\begin{eqnarray}
v_k(\eta_0)&=&\frac{1}{\sqrt{k}}\left[1+\frac{X+Y}{2}\theta_0
+{\cal{O}}(\theta_0^2)\right], 
\nonumber  \\
v'_k(\eta_0)&=&-i\sqrt{k}\left[1+\frac{Y-X}{2}\theta_0
+{\cal{O}}(\theta_0^2)\right ].
\label{Eqn:ModeInitial}
\end{eqnarray}
Here, $X$ and $Y$ are two complex parameters with the normalization condition
(\ref{Eqn:ModeNormalization}) constraining $\text{Re}(Y)$ to be zero.  The 
choice $X=Y=0$ corresponds to the initial
conditions for the conventional vacuum choice.

The parameter $\theta_0$ is defined as
\begin{equation}
\theta_0\equiv\left.\frac{aH}{k}\right |_{\eta_0},
\label{Eqn:ThetaParameter}
\end{equation}
which is the ratio of the physical size of a comoving mode $k$ to the Hubble
length at time $\eta_0$.  Note that $\theta_0$ vanishes in the limit 
when the physical size of the mode $k$, $\lambda=a/k$, is much smaller than 
the Hubble radius ($\lambda\ll
H^{-1}$), as well as when there is no expansion ($H\rightarrow 0$). In these
cases, the comoving mode $k$ does not see the curvature of the Universe, and 
we recover the Minkowski vacuum as a result.

For each mode, there is an ambiguity in the time $\eta_0$ when
$v_k(\eta_0)$ an $v_k'(\eta_0)$ are specified.  Conventionally, the
initial conditions for all of the modes are specified at
$\eta_0\rightarrow -\infty$.  However, that limit corresponds to a 
vacuum choice made when the physical length of the mode 
$(k/a)^{-1}\rightarrow 0$, where we expect conventional physics to 
break down.  Following Refs.~\cite{Danielsson02A,Easther02,Jacobson93}, it
seems more sensible to define the vacuum for each mode at the time
when the physical length of the mode equals a fixed length scale
$\Lambda^{-1}$, so that 
\begin{equation}
\frac{a(\eta_0(k))}{k}=\Lambda^{-1}. 
\label{Eqn:ModeCrossingTime}
\end{equation}
This introduces a cutoff scale $\Lambda$ into the theory, which can
be interpreted as the energy scale where the quantum field theoretic
description of the perturbations breaks down.  The cutoff scale is a free
parameter of the theory, although this is normally taken
to be the Planck scale.  With this prescription, the vacuum for each
mode $k$ is defined at a different time $\eta_0(k)$, and $\theta_0$ has
the following scaling relation:
\begin{equation}
\theta_0=\frac{H_*}{\Lambda}\left(\frac{k}{k_*}\right)^{-\epsilon},
\label{Eqn:ThetaScale}
\end{equation}
where $H_*$ is the Hubble parameter evaluated at the time the physical size of 
a comoving reference scale $k_*$ crosses the cutoff $\Lambda^{-1}$.

In Eq. (\ref{Eqn:ModeInitial}), $X$ and $Y$ can generally be functions of $k$. 
However, since they are dimensionless quantities, they can only depend on ratios 
of dimensionful quantities. Without introducing new scales into the theory, 
the only such ratios are $k/aH$ and higher time derivatives such
as $(k/a)'/H'$. In the former case, we can expand $X$ and $Y$ as 
functions of $\theta_0$ and recover the form of Eqs.
(\ref{Eqn:ModeInitial}). In the latter case, we expand $X$ and $Y$ as functions
 of $\theta_0$, the slow-roll parameters, and their time derivatives.  
Since we have fixed the slow-roll parameters as constants of the theory, they 
can similarly be absorbed into the parameterization without any loss of
generality. 

The solution to the equation of motion (\ref{Eqn:ModeEquationOfMotion}) is 
given by
\begin{equation}
v_k(\eta)=|\eta|^{1/2}[A_k H^{(1)}_{\nu}(|k\eta|)+B_kH^{(1)*}_{\nu}(|k\eta|)],
\label{Eqn:ModeSolution}
\end{equation}
where the Hankel function index is related to the slow-roll parameters by
\begin{equation}
\nu=\frac{3}{2}+2\epsilon-\delta. \label{Eqn:HankelIndex}
\end{equation}
The integration constants $A_k$ and $B_k$ are fixed by the initial
conditions (\ref{Eqn:ModeInitial}) to be
\begin{eqnarray}
A_k&=&\sqrt{\frac{\pi}{2}}\left[1+Y\frac{\theta_0}{2}
-i\left(\frac{2\nu+1}{4}\right)\theta_0\right]e^{i\sigma}, 
\label{Eqn:AkCoefficient} \\
B_k&=&\sqrt{\frac{\pi}{2}}X\frac{\theta_0}{2}e^{i\sigma}, 
\label{Eqn:BkCoefficient} \\
\sigma&=&\frac{-1}{\theta_0(1-\epsilon)}+\pi\left(\frac{\nu}{2}
+\frac{1}{4}\right). 
\label{Eqn:PhaseCoefficient}
\end{eqnarray}

Different choices of vacua leave different imprints upon the temperature
anisotropies in the cosmic microwave background radiation. For scalar
perturbations, the amplitude of the fluctuations can be characterized
by the power spectrum of the Bardeen variable
$\hat{\zeta}\equiv\hat{v}/z$, defined through the two-point correlation
function as
\begin{equation}
\langle 0|\hat{\zeta}(\mathbf{x},\eta)
\hat{\zeta}(\mathbf{x}+\mathbf{r},\eta)|0 \rangle \equiv 
\int \frac{dk}{k} \frac{\sin{kr}}{kr}\Pzeta(k). 
\label{Eqn:PowerSpectrumImplicit}
\end{equation}
Substituting Eq. (\ref{Eqn:ModeExpansion}) into
Eq. (\ref{Eqn:PowerSpectrumImplicit}) and using Eq. (\ref{Eqn:slowrollepsilon}), we 
find that
\begin{equation}
\Pzeta(k)=\frac{1}{8m_p^2\pi^2}\frac{k^3}{a^2\epsilon}|v_k|^2.
\label{Eqn:PowerSpectrumEquation}
\end{equation}
With these equations, 
we can compute the power spectrum for long wavelength modes
$|k\eta|\ll 1$, which is the spectrum of fluctuations that seed the
microwave background.  Substituting Eqs.
(\ref{Eqn:ModeSolution}), (\ref{Eqn:AkCoefficient}),
(\ref{Eqn:BkCoefficient}) and $n_s=1-4\epsilon+2\delta$ into Eq.
(\ref{Eqn:PowerSpectrumEquation}), we obtain the result
\begin{equation}
\Pzeta(k)=\Delta^2_0\left\{1+\theta_0|X|\cos
\left[\frac{2}{\theta_0(1-\epsilon)}+\varphi-2\pi\left(\frac{3-n_s}{4}\right)
\right]\right \}.
\label{Eqn:PowerSpectrum}
\end{equation}
Here, $\varphi$ is the phase of $X$, i.e., $X\equiv|X|e^{i\varphi}$, and 
$\Delta^2_0$ is the uncorrected power spectrum
\begin{equation}
\Delta^2_0=\delta_{\zeta}^2\left(\frac{k}{k_*}\right)^{n_s-1},
\label{Eqn:UnmodifiedPS}
\end{equation}
with $\delta_{\zeta}^2$ specifying the normalization of the power spectrum.
Note that $Y$ does not appear to first order.

The modification to the power spectrum (\ref{Eqn:PowerSpectrum}) consists of 
an oscillatory piece with a $k$-dependent amplitude.  A more convenient 
choice of parameters would be related directly to the characteristics of the
oscillation, rather than to the theoretical model.  To rewrite the power 
spectrum, we introduce the following parameters:
\begin{eqnarray}
\lambda&\equiv&|X|\frac{H_*}{\Lambda}, \label{Eqn:OscillationAmplitude} \\
\omega&\equiv&\frac{2\epsilon}{1-\epsilon}\frac{\Lambda}{H_*}, 
\label{Eqn:OscillationFrequency}\\
\alpha &\equiv& \varphi - 2\pi\left(\frac{3-n_s}{4}\right).
\label{Eqn:OscillationPhase}
\end{eqnarray}

The parameter $\lambda$ is the amplitude of the oscillation, modulo
the $k$ dependence $(k/k_*)^{-\epsilon}$, and $\omega$ is the
oscillation frequency in $\log (k/k_*)$.  Lastly, $\alpha$ characterizes the
phase of the oscillation.  Previous 
studies~\cite{Bergstrom02,Elgaroy03,Martin03} do not explicitly include the 
phase as a free parameter in their computation of constraints.  However, 
since the phase shifts
the starting point of the oscillations, this should not be neglected
in assessing the observability of the oscillations.

With these parameters, the power spectrum 
(\ref{Eqn:PowerSpectrum}) is rewritten in its final form 
\begin{equation}
\Pzeta(k)=\Delta^2_0\left \{1+
\lambda\left(\frac{k}{k_*}\right)^{-\epsilon}
\cos\left[ \frac{\omega}{\epsilon}\left(\frac{k}{k_*}\right)^{\epsilon}
+\alpha\right ]\right \}.
\label{Eqn:ReducedPowerSpectrum}
\end{equation}
The choice $\lambda=\zeta$, $\omega=2\epsilon/\zeta$, and $\alpha=-\pi/2$
corresponds to the two-parameter model considered in~\cite{Elgaroy03}, with
$\zeta,\epsilon\in[0,1]$.

The amplitude of the oscillation, $|X|\frac{H_*}{\Lambda}$, is
dependent on the parameters $|X|$ and $\Lambda$, but
the frequency $\frac{2\epsilon}{1-\epsilon}\frac{\Lambda}{H_*}$
depends only on $\Lambda$. Thus a measurement of the frequency
$\omega$ allows us to break the degeneracy in the amplitude.

Finally we comment on the possible range of values that
$|X|$ and $|Y|$ can take. Because Eqs. (\ref{Eqn:ModeInitial})
are expansions in $\theta_0$, there is no upper bound for
$|X|$ and $|Y|$ if we do not require the expansion to converge.
However, there are a couple of reasons why we should
expect $|X|$ and $|Y|$ to be small. The first is a theoretical prejudice: if 
$|X|$ and $|Y|$ are  not small, then there is no reason to expect the 
${\cal{O}}(\theta_0^2)$ and higher-order terms to be negligible.
In addition, limits on gravitational particle production
may constrain $|X|<5.5\frac{m_p}{\Lambda}$~\cite{ArmendarizLim03}. 

\subsection{Primordial Power Spectrum to the CMB}
\label{Subsect:CMBTheory}

The temperature and polarization anisotropies of the microwave background 
arise from acoustic oscillations of the inhomogeneous photon-baryon fluid,
which are seeded by the metric fluctuations created during inflation.  
The multipole moments
of the fluctuations can be related to the Bardeen variable $\zeta(\vc{k})$ as 
\begin{equation}
X_l^m = \frac{4\pi}{\sqrt{2l(l+1)}}(-i)^l\int\frac{d^3k}{(2\pi)^3}T_l^X(k)
Y_l^m{}^*(\vc{\hat{k}})\zeta(\vc{k}),
\label{Eqn:XlmZetaRelation}
\end{equation}
where $X=\{T,E\}$ is the temperature or $E$-mode multipole moments, and 
$T_l^X(k)$ denotes the transfer function encoding the acoustic physics and 
the projection effects (see, e.g., Ref.~\cite{HuOkamoto03} for discussion on 
the structure of the transfer functions).  

The CMB power spectra, defined as
\begin{equation}
\avg{X_l^m{}^*X'{}_{l'}^{m'}}\equiv C_l^{XX'}\delta_{ll'}\delta_{mm'},
\label{Eqn:ClDefinition}
\end{equation}
are then related to the primordial power spectrum $\Delta_\zeta^2(k)$ by
\begin{equation}
\frac{l(l+1)}{2\pi}C_l^{XX'} = \int\frac{dk}{k}T^X_l(k)T^{X'}_l(k)\Pzeta(k).
\label{Eqn:CltoPzeta}
\end{equation}

The transfer functions in Eq. (\ref{Eqn:CltoPzeta}) limit our ability to 
measure $\Pzeta(k)$ in several ways.  First, $T^X_l(k)$ projects
the fluctuations onto a sphere, and this projection peaks along 
$l_{\text{peak}}\approx kr_*$ 
with a width of $\Delta l\approx 20$, where $r_*=13.8 \text{Gpc}$ is the 
angular
diameter distance to the last scattering surface.  Since a Fourier 
mode with wave number $k$ projected onto a sphere also contributes to 
multipole moments with $l\ll l_{\text{peak}}$, power at wave number $k$ gets
distributed to multipole moments with $l\ll l_{\text{peak}}$, as well as in
the narrow band with $\Delta l\approx 20$ around  $l\ll l_{\text{peak}}$.  

The transfer functions $T_k^X(k)$ also include contributions that arise due
to secondary effects.  In the low multipole regime, the temperature
transfer function includes a contribution from the integrated
Sachs-Wolfe effect~\cite{ISW}, which provides additional power at
lower multipoles.  Rescattering of the CMB photons during reionization
contributes to a feature in the low multipole regime of the
polarization transfer function. This shows up as a bump in the $EE$ power
spectrum and $TE$ cross spectrum, of which the latter was first
detected by the WMAP collaboration~\cite{Kogut03}.  Both 
features tends to broaden features in the primordial power
spectrum.  Although there are other foreground effects at high
multipoles, we work in the multipole regime below which those
secondaries are insignificant. 

An additional limitation arises due to gravitational lensing of the
CMB~\cite{Seljak96,Zaldarriaga98,Hu00}.  Lensing has the effect of
redistributing power between adjacent multipole bins, thereby reducing
the width of resolvable features to roughly $\Delta l\sim 60$.  This
translates to a maximum log frequency of roughly $\omega\lesssim 30$,
and potentially an important limitation on measuring the properties of
the primordial fluctuations~\cite{HuOkamoto03}.  We will restrict the
range of $\omega$ to be below 30 in our analysis to avoid the regime
where contamination due to lensing is an important effect.

\section{Analysis}
\label{Sect:Analysis}

We are interested in how well the microwave background can in
principle distinguish signatures of new physics from the standard
inflationary models.  Previous works have estimated the uncertainties
on parameters related to the cutoff physics using the Fisher matrix
formalism~\cite{Bergstrom02}, or used a likelihood approach on a
fiducial model containing a modulation~\cite{Elgaroy03}.  Both
start from a two-parameter model for the primordial power
spectrum~\cite{Danielsson02A}, and therefore do not incorporate the
full range of effects that might arise from the existence of a high-energy 
cutoff.  Furthermore, the analyses of the ideal case depend on
the choice of fiducial models.  We currently have no \textit{a priori}
knowledge, either from experiments or theoretical prejudice, about
what set of parameters would constitute a reasonable fiducial model other
than the standard inflationary model without oscillations.  However, Refs.
~\cite{Bergstrom02} and ~\cite{Elgaroy03} both assume fiducial models 
containing oscillations, and the resulting parameter forecasts therefore 
only apply to those fiducial models.

In this work, we instead present two analyses, both of which
complement the existing literature.  In the first, we use current CMB
data sets to place constraints on the four-parameter description of the 
oscillation induced by the phenomenology at the cutoff scale.  
In the second, we present the fundamental 
limitations on measurements of the oscillation parameters that arise from the 
statistics of CMB measurements.

Both of these analyses rely on the calculation of the likelihood over
the expanded parameter set, which is accomplished using Markov Chain
Monte Carlo (MCMC) methods.  In the following sections, we briefly
describe the calculation of likelihoods for microwave background data,
as well as details concerning our use of MCMC.

\subsection{Likelihood Analysis}
\label{Subsect:likelihood}

Let us suppose that we are given a data set $\hat{C}_l$, along with a set of 
model power spectra $C_l(\vc{p})$, with $\vc{p}$ denoting a vector in an 
$N$-dimensional parameter space.  The posterior probability distribution 
$P(\vc{p}|\hat{C}_l)$, which is the probability density at point 
$\vc{p}$ in the parameter space given the data set, is then computed by
\begin{equation}
P(\vc{p}|\hat{C}_l)\propto P(\vc{p})\like(\hat{C}_l|\vc{p}).
\label{Eqn:Posterior}
\end{equation}
Here, $P(\vc{p})$ represents our prior assumptions about the parameters,
and $\like$ denotes the likelihood function.  The proportionality constant
is set by requiring that
\begin{equation}
\int d\vc{p} P(\vc{p}|\hat{C}_l) = 1
\label{Eqn:PosteriorNorm}
\end{equation}
over the allowed range in parameter space.  In many instances, we do not have
any prior knowledge about what the parameter should be, so that we assign a 
flat prior $P(p^i)=1$ for the $i$th parameter.  In such a case, the posterior 
probability is proportional to the likelihood function, so maximizing the 
likelihood function in the parameters is equivalent to finding the point with 
the maximum probability density.  To find the 1$\sigma$ uncertainties around 
the maximum likelihood point, we then find contours that enclose 68.3\% of the 
posterior probability.

Assuming that the CMB multipole moments are Gaussian distributed, the 
likelihood for an all-sky, noiseless experiment is given by
\begin{equation}
\like \propto \prod_{lm}\frac{1}{\sqrt{\det{\mat{C}_l}}}
\exp\left \{-\frac{1}{2}\vc{d}^\dag{}_l^m \mat{C}^{-1}_l\vc{d}_l^m\right \},
\label{Eqn:Like}
\end{equation}
where the data vector is given by $\vc{d}_l^m \equiv \{T_l^m,E_l^m \}$,
and the covariance matrix is
\begin{equation}
\mat{C}_l = \left (\begin{array}{cc}
C_l^{TT} & C_l^{TE} \\
C_l^{TE} & C_l^{EE}
\end{array}\right ),
\label{Eqn:XlmCovariance}
\end{equation}
with $C_l^{XX'}$ denoting the theoretical power spectra.
Defining the estimator for the power spectra according to
\begin{equation}
\hat{C}_l^{XX'} \equiv \sum_m\frac{|X^*{}_l^mX'{}_l^m|}{2l+1},
\end{equation}
the log likelihood function becomes (up to an irrelevant constant)
\begin{equation}
-2\ln\like = \sum_l(2l+1)\left [
\ln\left (\frac{C_l^{TT}C_l^{EE}-(C_l^{TE})^2}
{\hat{C}_l^{TT}\hat{C}_l^{EE}-(\hat{C}_l^{TE})^2}\right )
+ \frac{\hat{C}_l^{TT}C_l^{EE}+\hat{C}_l^{EE}C_l^{TT}-2\hat{C}_l^{TE}C_l^{TE}}
{C_l^{TT}C_l^{EE}-(C_l^{TE})^2}- 2 \right ].
\label{Eqn:LogLike}
\end{equation}
We have chosen the constants so that $-2\ln\like=0$ when the theoretical
power spectra $C_l^{XX'}$ are equal to the data estimates $\hat{C}_l^{XX'}$,
such that $-2\ln\like$ behaves as an effective chi-squared statistic.  We will
use this form of the likelihood function in our analysis of the idealized case.

If only the temperature power spectrum is measured, then the log likelihood 
function is given by
\begin{equation}
-2\ln\like^{TT} = \sum_l(2l+1)\left [
\ln\left (\frac{C_l^{TT}}{\hat{C}_l^{TT}}\right )
+ \frac{\hat{C}_l^{TT}}{C_l^{TT}}- 1 \right ].
\label{Eqn:LogLikeTT}
\end{equation}
For a realistic experiment, the presence of sky cuts and instrumental noise
necessitate the use of various approximations to the likelihoods, 
Eqs. (\ref{Eqn:LogLike}) and (\ref{Eqn:LogLikeTT}).    

The transfer functions $T_l^X(k)$ used to calculate the model power
spectra $C_l^{XX'}$ can be obtained by using publicly available
Boltzmann integral codes such as CMBFAST~\cite{CMBFAST} and
CAMB~\cite{CAMB}.  However, these codes sample the transfer function
sparsely in $k$ and $l$ to reduce computation time, and the resulting
power spectra can deviate from the true power spectra if there are
features in the initial power spectrum which have scales comparable to
the gridding.  In our parametrization, higher values of $\omega$
correspond to higher oscillation frequencies, so we expect the
computation from fast Boltzmann codes to become less reliable in that
regime.

\begin{figure}[tb]
\myfigure{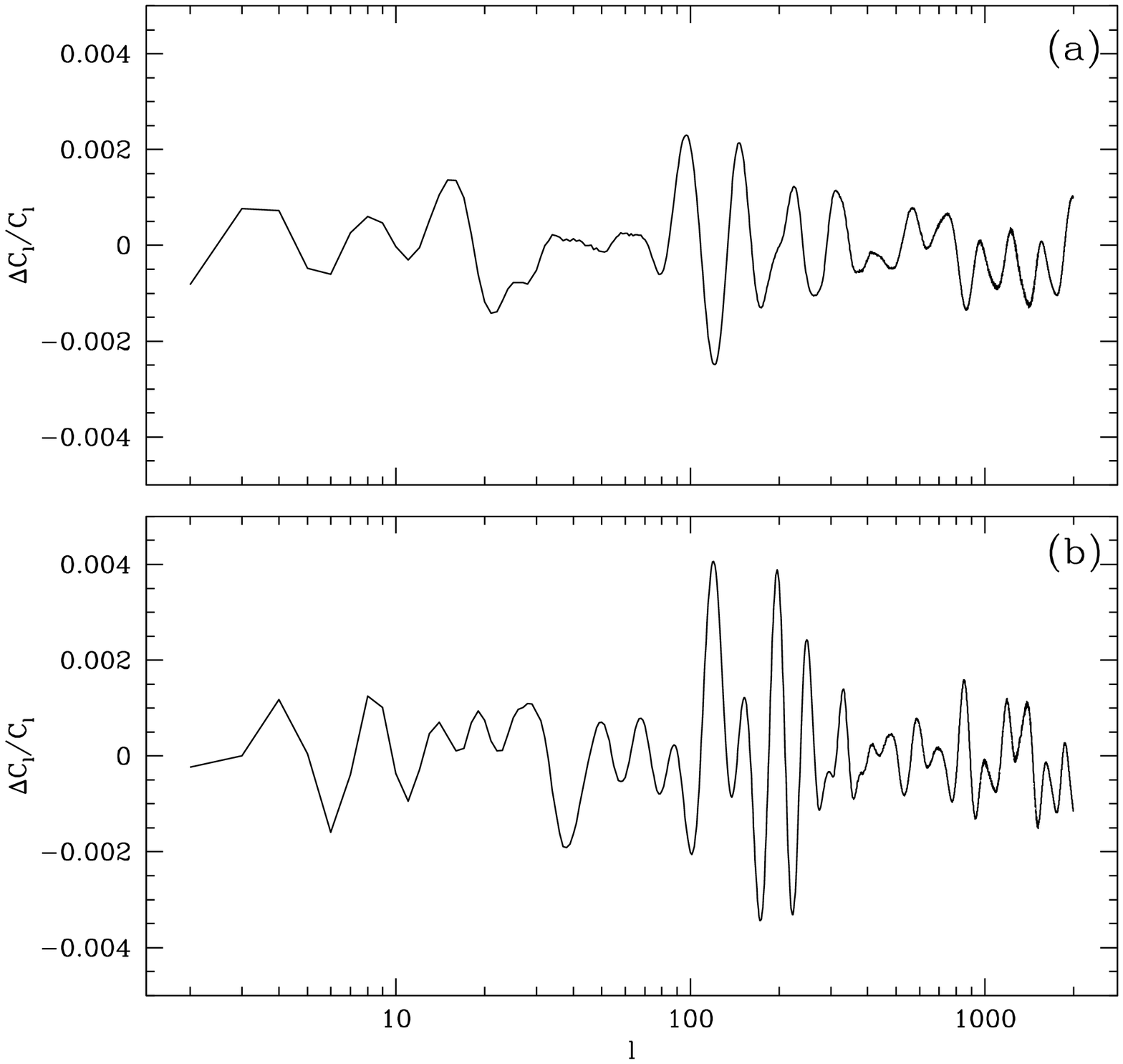}
\caption{Fractional deviation of CAMB computation from the true power spectrum,
for (a) $\lambda=0.01$, $\epsilon = 0.2$, $\omega=20$, $\alpha=0$, and (b)
$\lambda=0.01$, $\epsilon = 0.2$, $\omega=30$, $\alpha=0$.  At $\omega=20$, 
the deviations are at around $0.2\%$, while at $\omega=30$, the deviations 
approach $0.4\%$.}
\label{Fig:compareCAMB}
\end{figure}

To illustrate this, we compare the power spectra 
generated using CAMB to those from a Boltzmann hierarchy code developed by 
Hu~\cite{HuOkamoto03}, which has the advantage of capturing the high 
frequency oscillations that might be problematic with the integral codes.
To determine the frequency limit imposed by the accuracy of the code, we 
compare the power spectra generated by CAMB to that from the hierarchy code
for various values of $\omega$.  As can be seen in Fig.~\ref{Fig:compareCAMB},
where we plot the fractional deviation
\begin{equation}
\frac{\Delta C_l}{C_l} \equiv \frac{C_l^{\text{CAMB}}
-C_l^{\text{true}}}{C_l^{\text{true}}},
\label{Eqn:FracDev}
\end{equation} 
the amplitude of the deviation increases with $\omega$.  We impose the
fairly strict condition that deviations be less than $0.2\%$.  As a
result, the range of $\omega$ is restricted to be less than 20 for
computations performed using CAMB.

Since computing the likelihood function from $T^X_l(k)$ on a grid over the 
entire parameter space is generally not feasible for more than two parameters, 
we utilize Markov Chain Monte Carlo sampling to compute the posterior 
probability distribution.

\subsection{Markov Chain Monte Carlo}
\label{SubSect:MCMC}

Markov Chain Monte Carlo (MCMC) methods have been applied to CMB
analysis of the data from several experiments~\cite{LewisBridle02,Verde03}. 
We give a brief overview of the method, as well as a description of our
implementation, and refer the reader to
Refs.~\cite{GilksBook,LewisBridle02} for full treatments.

MCMC is a method designed to simulate the posterior probability distribution 
through a series of random draws.  In our case, the MCMC produces a set of 
$N$ points $\{\vc{p}_t,t=1\ldots N\}$ that fairly sample the distribution 
$P(\vc{p}|\hat{C}_l)$, so that the density of points $\{\vc{p}_t\}$ is 
proportional to the posterior distribution.  Statistical properties of the 
parameters $p^i$ can then be estimated using the corresponding properties of 
the samples; for instance, the mean and variance of 
parameter $p^i$ can be estimated by using the sample mean and variance via
\begin{eqnarray}
\bar{p}^i & = & \frac{1}{N}\sum_{t=1}^N p^i_t, \nonumber \\
\sigma^2(p^i) & = & \frac{1}{N-1}\sum_{t=1}^N (p^i_t - \bar{p}^i)^2.
\end{eqnarray}

Our goal is to produce a chain whose stationary distribution is the
distribution $P(\vc{p}|\hat{C}_l)$.  We describe the
Metropolis-Hastings algorithm~\cite{Metropolis53,Hastings70}, used in
many MCMC implementations for CMB analysis including the WMAP
analysis~\cite{Verde03} and the public package
CosmoMC~\cite{LewisBridle02}, as well as in our own implementation.
We refer the reader to Ref.~\cite{LewisBridle02} for details on
CosmoMC, which we utilize in Sec.~\ref{Sect:WMAPBounds}, and give a
generic overview of the algorithm here.

  At each step $t$ in this algorithm, a 
candidate point $\vc{q}$ is first chosen from a proposal distribution 
$\beta(\vc{q}|\vc{p}_t)$.  The point $\vc{q}$ is accepted with 
probability
\begin{equation}
\pi(\vc{p}_t,\vc{q}) = \min\left (1,
\frac{P(\vc{q}|\hat{C}_l)\beta(\vc{p}_t|\vc{q})}
{P(\vc{p}_t|\hat{C}_l)\beta(\vc{q}|\vc{p}_t)}\right ).
\end{equation}
If $\vc{q}$ is accepted, then the chain continues, with
$\vc{p}_{t+1}=\vc{q}$; otherwise, $\vc{p}_{t+1} = \vc{p}_t$, i.e., the
chain does not move.  Note that the proposal distribution can depend
on the last point of the chain.  We take $\beta(\vc{q}|\vc{p}_t)$ to
be a Gaussian distribution with mean $\vc{p}_t$ and a fixed, diagonal
covariance matrix.  We first run a short Markov chain to determine the
variance to be used in the proposal distribution.

Proper conclusions from chains are not possible without determining if the 
chains have converged to the stationary distribution.  The first step in 
ensuring this is to remove the first few thousand points belonging to a 
burn-in period.  All points past the burn-in period are assumed to have 
forgotten about its starting point, and can be thought of as being sampled from
the stationary distribution.

Additionally, we employ a convergence criterion, advocated by Gelman and 
Rubin~\cite{GelmanRubin92} and used by the WMAP 
collaboration~\cite{Verde03,Spergel03a,Peiris03}, to ensure proper mixing and convergence of 
chains.  In this criterion, one generates several chains from different 
starting points, and quantifies whether these chains are indistinguishable from
each other.  

Suppose one generates $M$ chains from different starting points, each chain 
with $2N$ points.  For each parameter $p^i$, we take the last $N$ points from
each chain, and label the sequences as $\{p^i_{\alpha\beta}\}$, with 
$\alpha=1,\ldots,N$, and $\beta=1,\ldots,M$.  Let the mean of each chain be
defined by
\begin{equation}
\bar{p}^i_\beta = \frac{1}{N}\sum_{\alpha=1}^{N} p^i_{\alpha\beta},
\label{Eqn:ChainMean}
\end{equation}
and the mean of the distribution by
\begin{equation}
\bar{p}^i = \frac{1}{MN}\sum_{\alpha=1}^N\sum_{\beta=1}^Mp^i_{\alpha\beta}.
\label{Eqn:DistroMean}
\end{equation}
We then define the variance between chains $B^i$ and the variance within
chains $W^i$ as
\begin{eqnarray}
B^i &=& \frac{N}{M-1}\sum_{\beta=1}^M (\bar{p}^i_\beta - \bar{p}^i)^2,
\label{Eqn:VarBetChains}\\
W^i &=& \frac{1}{M(N-1)}\sum_{\beta=1}^M\sum_{\alpha=1}^N
(p^i_{\alpha\beta}-\bar{p}^i_\beta)^2.
\label{Eqn:VarInChains}
\end{eqnarray}
Lastly, we can define an estimate of the variance of $p^i$ using
\begin{equation}
V^i = \frac{N-1}{N}W^i + \frac{1}{N}B^i.
\label{Eqn:VarEstimate}
\end{equation}

The estimate $V^i$ is an unbiased estimator of the variance if the 
chains are stationary, but overestimates the variance if the starting points
of the chains are dispersed.  On the other hand, $W^i$ typically underestimates
the variance, since the finite chains would not have had enough steps to 
range over the entire distribution.  Both $V^i$ and $W^i$ approach 
the true variance in the limit $N\rightarrow\infty$, but from opposite sides.
If we now form the ratio
\begin{equation}
R^i \equiv \frac{V^i}{W^i},
\label{Eqn:GelmanRubinR}
\end{equation}
this ratio should approach $1$ from above as the chains grow long, i.e., as 
the chains more faithfully reproduces the target distribution, and small
values of $R^i$ indicate that the chains have mixed well and converged.
We take the condition $R^i < 1.1$ for all parameters as our convergence 
criterion.

\section{Current Bounds from CMB Data}
\label{Sect:WMAPBounds}

We first present the constraints on the 
oscillation parameters using the WMAPext data set~\cite{Spergel03a}, 
which consists of the WMAP 1-year data set~\cite{Kogut03}
supplemented by the Arcminute Bolometer Array Receiver(ACBAR)~\cite{ACBAR} and 
Cosmic Background Imager (CBI)~\cite{CBI} data sets to
extend the multipole range.  Although this analysis is similar to
those presented in ~\cite{Bergstrom02,Elgaroy03,Martin03}, our
parametrization of the primordial power spectrum removes some of the
degeneracies present in the original parametrization.

In our analysis, we focus on the constraints that can be placed on the
parameters defining the oscillations in the presence of cosmological
parameters.  We vary the following four cosmological parameters
($\Omega_bh^2,\Omega_{\text{CDM}}h^2,H_0,z_{\text{re}}$), as well as
the six parameters ($A_s,n_s,\lambda,\epsilon,\omega,\alpha$) that
define the primordial power spectrum, assuming flatness of the
universe.  Here, $A_s$ parametrizes the normalization 
of the primordial power spectrum according to 
\begin{equation}
\delta_\zeta^2 = 10^{-10}A_s,
\label{Eqn:CAMBNorm}
\end{equation}
and matches the normalization convention used in CAMB.

The likelihood contours are calculated using 
CosmoMC,
with the power spectra computed from a version of CAMB modified to
incorporate the oscillatory power spectrum.  We assume a flat prior
for all parameters, restricting the ranges of $\omega$ and $\epsilon$
to be $\omega<20$ and $\epsilon<0.2$, respectively. 

Our results are based on eight chains started at random points in the
10-dimensional parameter space.  We find that the chains have
converged after 100000 points, using the convergence criterion
presented in Sec.~\ref{SubSect:MCMC}.

\begin{figure}[tbhp]
\includegraphics{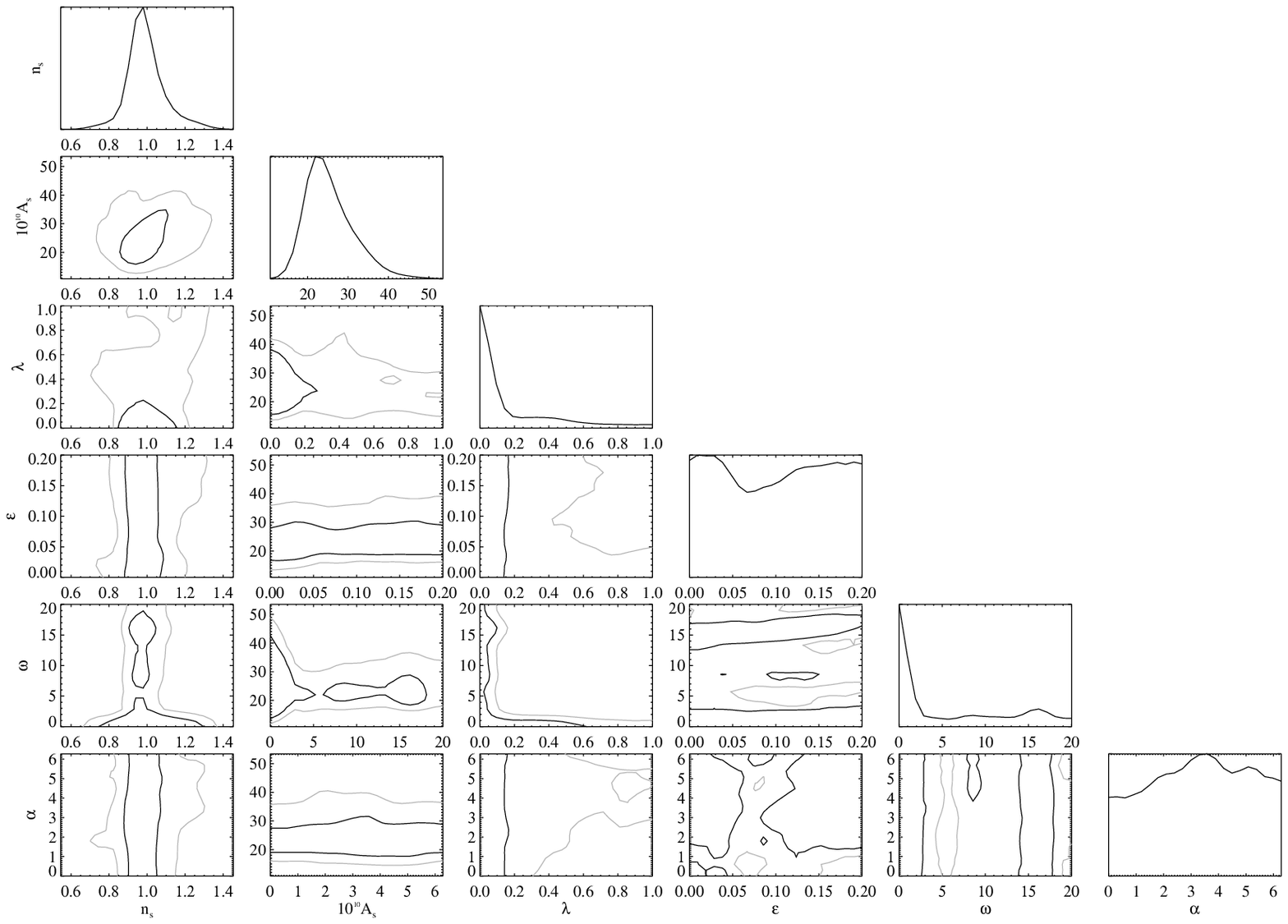}
\caption{WMAP likelihood contours for parameters 
describing the primordial power spectrum.  The diagonals show the marginalized
likelihoods as functions of the individual parameters.  The off-diagonals show
the two-dimensional likelihoods after marginalizing over the other parameters.
Solid lines denote the 1$\sigma$ contours, and the gray lines the 2$\sigma$
contours.}
\label{Fig:SixParsWMAP}
\end{figure}
\begin{table}[tbhp]
\begin{tabular}{l|cc}
Parameter & 1$\sigma$& 2$\sigma$ \\
\hline
$\lambda$ & $<0.20$ & $<0.77$ \\
$\epsilon$ & -- & -- \\
$\omega$ & $<10.1$& $< 17.8$\\
$\alpha$ & -- & --
\end{tabular}
\caption{Constraints on $\Pzeta(k)$ parameters from the 
WMAPext data set, indicating the 1$\sigma$ and 2$\sigma$ uncertainties on
parameters.  Dashes indicate that no meaningful constraints were possible.}
\label{Table:WMAPlimits}
\end{table}

The limits on the oscillation parameters are given in
Table~\ref{Table:WMAPlimits}, with the corresponding likelihood contours
shown in Fig.~\ref{Fig:SixParsWMAP}.  Not surprisingly, the current data set
does not give a strong constraint.  At 2$\sigma$, the amplitude of the 
oscillations are limited to $\lambda<0.77$, indicating consistency with a 
scale-invariant power spectrum.  Assuming $|X|=1$, this translates to 
a restriction of the cutoff scale to $\Lambda > 1.3 H_{\text{infl}}$,
and does not lead to any conclusions if we allow $|X|<1$.  

\begin{figure}[tbhp]
\myfigure{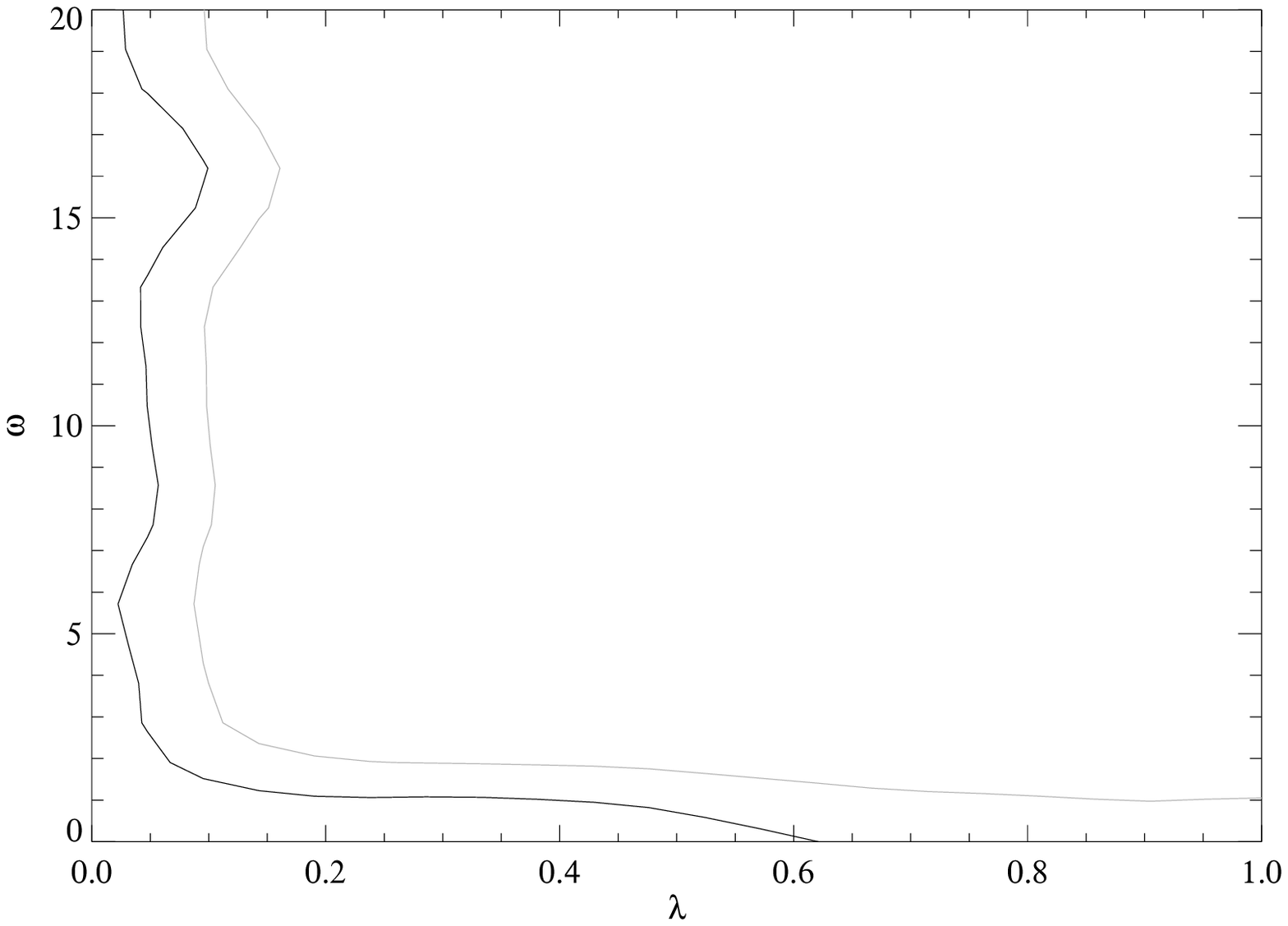}
\caption{Marginalized likelihood plot for the parameters $\lambda$ and
$\omega$, with the solid curve denoting the 1$\sigma$ contour and the gray
curve the 2$\sigma$ contour.  The contours curve around the $\lambda=0$ and 
$\omega=0$ lines and exhibit the degeneracy between $\lambda=0$ and $\omega=0$ 
models.}
\label{Fig:WMAP35}
\end{figure}

The lack of a strong constraint on $\lambda$ is largely due to a large
degeneracy between three parameters for a nonoscillatory fiducial
model.  Specifically, the three subspaces with $\lambda\rightarrow 0$,
$\omega\rightarrow 0$, and $\epsilon\rightarrow 0$ all correspond to
lack of oscillations, so that a data set consistent with a lack of
oscillations will have flat likelihoods along these planes.  This
behavior can be seen in the likelihood contours of
Fig.~\ref{Fig:WMAP35}, where we have marginalized the likelihood over
all parameters except $\lambda$ and $\omega$.  As can be seen, the
contours curve along the $\lambda=0$ and $\omega=0$ lines, indicating
that both $\lambda=0$ and $\omega=0$ are supported, so that strong
constraints cannot be placed on either of the two parameters.

Previous constraints using the current data set come from 
Refs.~\cite{Elgaroy03,Martin03}.  
Reference~\cite{Elgaroy03} finds that the current data set
is consistent with a scale-invariant power spectrum, only excluding the 
the region with $\lambda > 0.1$ and $0.1\le\epsilon\le 1$.  Since our results
indicate that, for the $\lambda$-$\epsilon$ plane, the exclusion region is 
given by $\lambda>0.5$, $\epsilon>0.05$ to 2$\sigma$, the two analyses are 
roughly in agreement.

In contrast, Ref.~\cite{Martin03} finds that a model with ($\lambda=0.29$, 
$\omega=288$, $\epsilon=0.02$) has an effective $\chi^2$ that is lower 
than the standard model without oscillations, with a greater than 3$\sigma$ 
significance.  We deliberately avoid the high-frequency regime, since 
gravitational lensing tends to suppress the oscillatory features for models 
with $\omega>30$.  We therefore cannot make any statements regarding the 
consistency of the results\footnote{See, however, \texttt{astro-ph/0402609}.}.

\begin{figure}[tbhp]
\myfigure{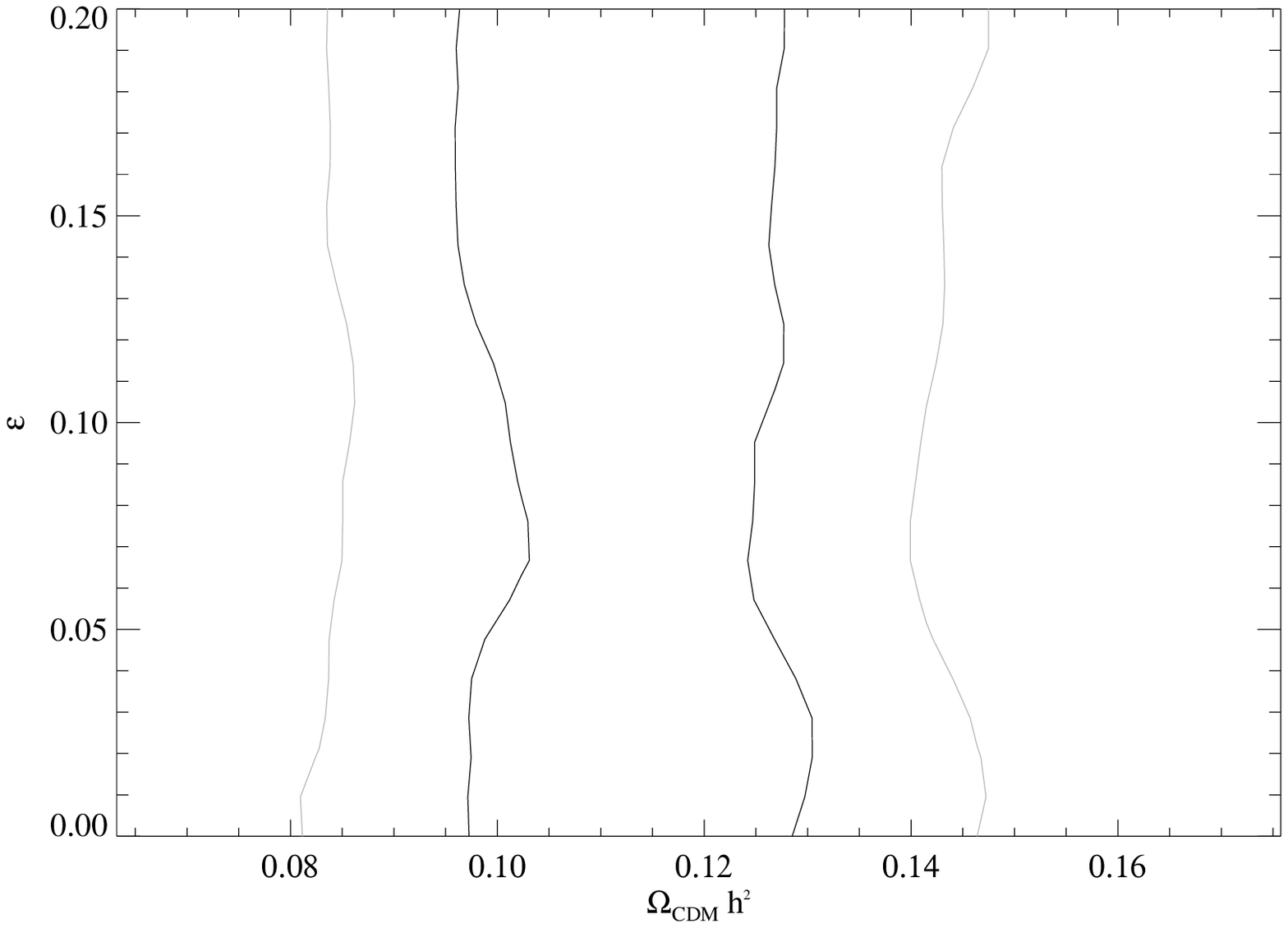}
\caption{Likelihood contours for $\epsilon$ vs. $\Omega_{\text{CDM}}h^2$.
There are no measurable correlations between $\epsilon$ and 
$\Omega_{\text{CDM}}h^2$.} 
\label{Fig:CosmoExample}
\end{figure}
Lastly, we comment on the relation between the cosmological parameters and
the oscillation parameters.  As expected, we find that the cosmological 
parameters are essentially 
independent from the oscillation parameters.  The cosmological parameters 
(e.g., $\Omega_bh^2$) influence the structure of the acoustic peaks, which 
have roughly equal spacings in $k$.  On the other hand, the oscillations we 
consider are in $\log (k/k_*)$, and intrinsically vary separately from the 
acoustic peaks.  As an illustration, we show the likelihood contours for 
$\Omega_{\text{CDM}}h^2$ and $\epsilon$ in Fig.~\ref{Fig:CosmoExample}.
Thus, the additional parameters do not bias the measurements of the 
cosmological parameters.  

\section{Theoretical Limit}
\label{Sect:TheoryLimit}

We are ultimately interested in the intrinsic limitations of the CMB
as a probe of primordial physics.  To investigate this, we now consider an
experiment that is noiseless and measures the multipole moments out to
some maximum multipole $l_{\text{max}}$, so that the uncertainties in
the measurement are due to cosmic variance alone.  With this idealized
experiment, we use the transfer functions from the hierarchy code and
explore the six-dimensional parameter space defining the primordial
power spectrum with high precision.  Due to the computational expense
required, we choose to keep the cosmological parameters fixed.  While 
such a situation may be unrealistic, the resulting limit on our ability to
measure the parameters of the primordial power spectrum is robust, in
that any consideration of realistic effects can only make the
uncertainties larger.  Furthermore, since the results from the previous 
section indicate that the cosmological parameters are independent of the 
oscillatory parameters to $\omega_{\text{max}}=20$, fixing the cosmological 
parameters should not bias
our results.  Thus, the limitations on parameters we quote
should be considered as theoretical limits on the detectability of
cutoff physics, arising purely from the statistical properties of the
microwave background and the processing due to the radiation transfer
functions.

The maximum multipole $l_{\text{max}}$ of the idealized experiment is
set by considering the additional complications due to gravitational lensing 
and other secondary effects.  The power from
gravitational lensing becomes comparable to that from the primary
anisotropies at $l\approx 3000$, and presents a fundamental limit to
the range of multipoles we can use \cite{Hu00}.  However, other secondary 
effects, such as the thermal and kinetic Sunyaev-Z'eldovich 
effect~\cite{SZ}, become dominant by $l\sim 3000$.  Furthermore, 
deep field observations by CBI suggest an excess of power in the range 
$l\approx 2000-3500$, which may be due to the aforementioned secondary 
effects~\cite{CBIexcess}.  
We therefore choose a conservative value of $l_{\text{max}}=2000$ to
avoid the regime where secondary effects may be dominant. 

For this ideal case, we simulate a data set by generating a realization of 
CMB power spectra that corresponds to a primordial power spectrum with
$\delta_\zeta=5.07\times 10^{-5}, n_s=0.99$, and without the oscillatory 
features. For convenience, we reparametrize the normalization to
\begin{equation}
\delta_\zeta^2 = (5.07\times 10^{-5})^2A_f,
\label{Eqn:AsDefinition}
\end{equation}
so that $A_f=1$ corresponds to the fiducial model.  This is related to 
the normalization $A_s$ by $A_f = 3.89\times 10^{-2} A_s$, and is chosen 
for computational convenience.  The cosmological parameters
are fixed to be $\{\Omega_bh^2=0.024,\Omega_mh^2=0.1399,\Omega_{DE}=0.73,
\tau=0.17,T/S=0,w_{\text{DE}}=-1\}$ for both the simulated data set and the 
model power
spectra.  From a converged run consisting of three chains with 230000
points per chain,
we find the maximum likelihood point and uncertainties on the 
six parameters $\{A_f,n_s,\lambda,\epsilon,\omega,\alpha\}$ that specify the 
primordial power spectrum, where we impose the priors $\epsilon < 0.2$ 
and $\omega < 30$ to avoid the regime where gravitational lensing becomes
important.  Gravitational waves are assumed to be absent.

Since we have taken the oscillations to be absent in our simulated data set,
the resulting constraints on the oscillation parameters can be regarded as 
the extent to which the oscillatory parameters can deviate from the fiducial
model before being statistically different from the standard case.  For 
instance, a constraint of $\lambda < 0.1$ at 1$\sigma$ would imply that models
with $\lambda < 0.1$ are statistically indistinguishable from the standard 
model with $\lambda = 0$.  Such a bound can be thought of as the fundamental
lower limit achievable by a CMB experiment.   

\begin{figure}[tbhp]
\includegraphics{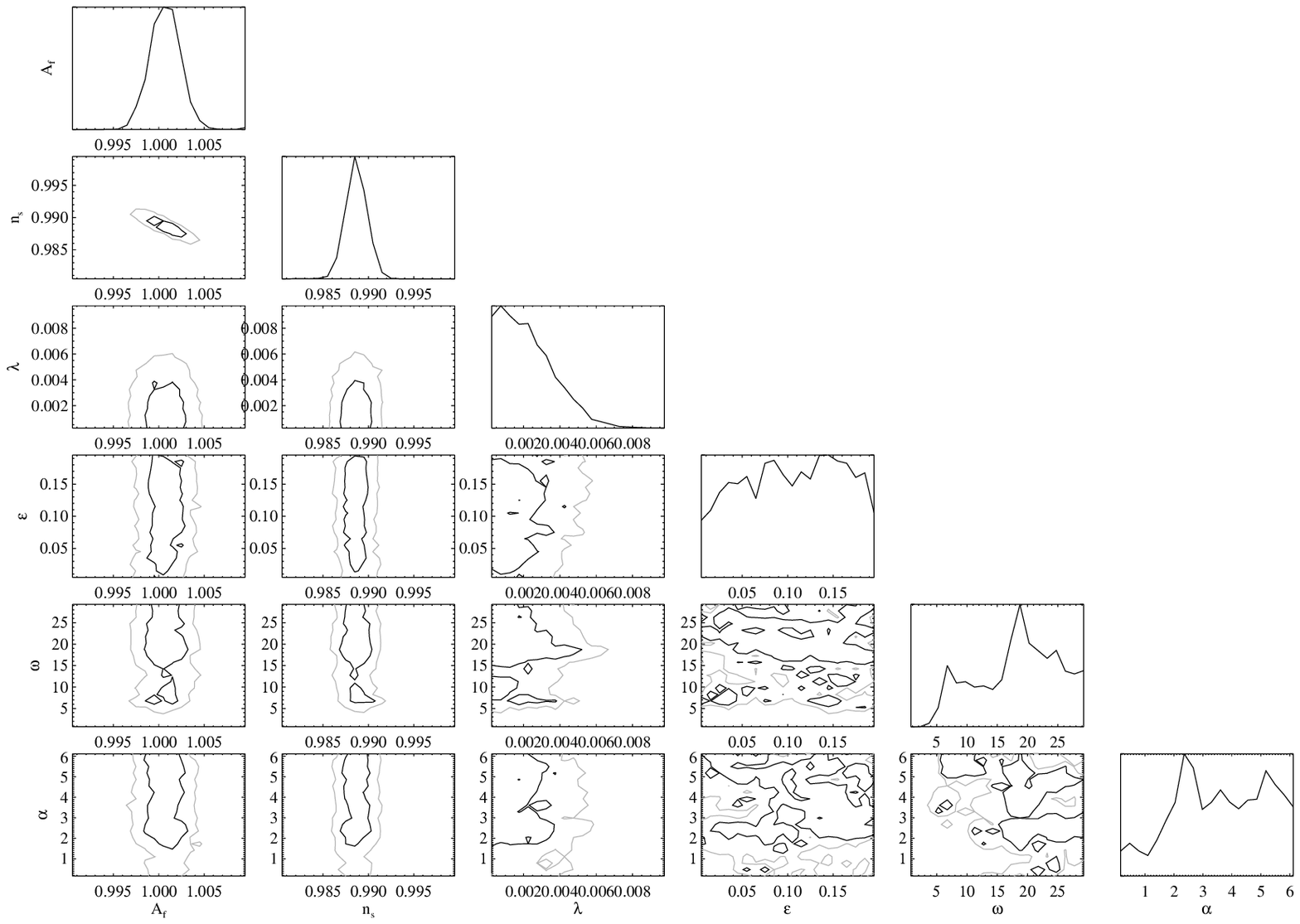}
\caption{1$\sigma$ and 2$\sigma$ likelihood contours from a cosmic variance
limited experiment, with a simulated data set containing no oscillations.  
The gray curved denote 2$\sigma$ contours.  The diagonals show the 
marginalized likelihoods as functions of the individual parameters.}
\label{Fig:SixParsCV}
\end{figure}

\begin{table}[bthp]
\begin{tabular}{l|ccc}
Parameter & mean & 1$\sigma$& 2$\sigma$ \\
\hline
$\lambda$ & $0.002$ & $\lambda<0.003$ & $\lambda<0.005$ \\
$\epsilon$ & $0.10$ & $0.07 < \epsilon < 0.14$ & $0.002 <\epsilon <0.19$\\
$\omega$ & $18.5$ & $16.1 < \omega <22.3 $& $ 6.8 < \omega < 28.6$\\
$\alpha$ & $3.6$ & $ 2.6 < \alpha < 4.7 $ & $ 0.7 < \alpha < 6.0 $
\end{tabular}
\caption{Constraints on $\Pzeta(k)$ parameters for a cosmic variance limited
experiment, assuming a fiducial model without oscillations.  In particular,
the amplitude of oscillations can be constrained to be below $0.5\%$.}
\label{Table:CVparams}
\end{table}
We present the limits in Table~\ref{Table:CVparams} and the likelihood
contours in Fig.~\ref{Fig:SixParsCV}, showing that a cosmic
variance limited experiment to $l\approx 2000$ can limit the oscillations to  
an amplitude of less than $0.5\%$ and frequencies of $\omega > 6.5$,
although other parameters are less constrained.  
As noted in Ref.~\cite{Elgaroy03}, the likelihood function has many local maxima
and irregularities.  Our parametrization seems to 
reduce the irregularities in the amplitude parameter $\lambda$; however,
features are evident in the two-dimensional likelihood contours of the other 
parameters.  While we believe our chains have converged, 
the presence of such features in the contour plots suggest that robust 
predictions for parameter uncertainties may require calculation of the 
likelihood function on much finer intervals.

\begin{figure}[ptbh]
\myfigure{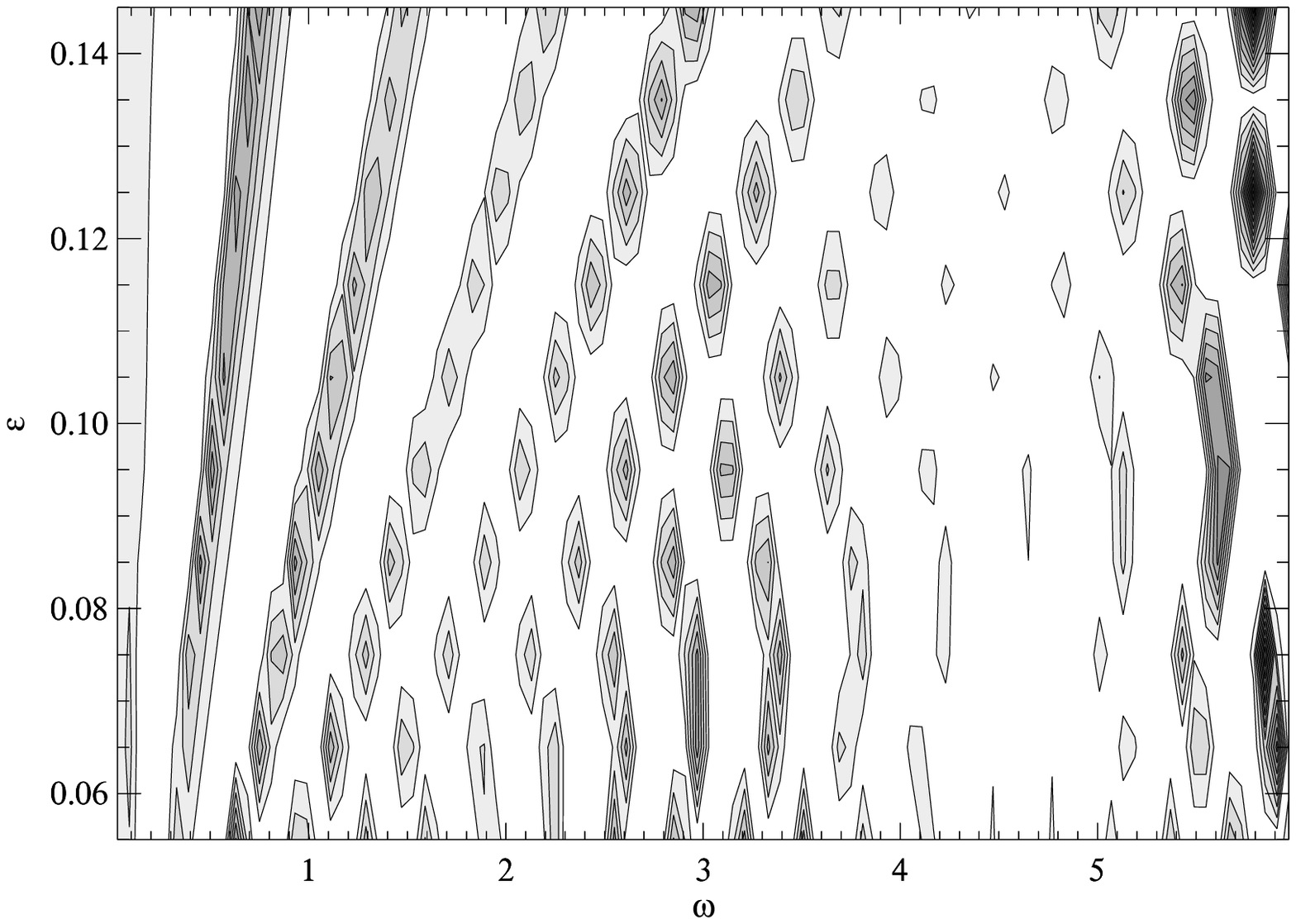}
\caption{The likelihood function, computed on a grid of $\omega$ and 
$\epsilon$, holding the other variables fixed at $\lambda=0.002$, $\alpha = 0$.
The curves are iso-likelihood contours, the levels evenly spaced between
minimum and maximum, with the darker spots having higher likelihood.
Near $\omega=5$, the local 
maxima seem to be lower than elsewhere, possibly accounting for the fact that 
the MCMC rules out $\omega < 6$.}
\label{Fig:slice}
\end{figure}

In particular, a plot of the likelihood function over
$\epsilon$ and $\omega$, with fixed amplitude and phase, reveals that the 
local maxima are distributed along lines of constant $\omega/\epsilon$, and
become sparse between $\omega=4$ and $\omega=5$ (Fig.~\ref{Fig:slice}).
The distribution along constant $\omega/\epsilon$ can be understood as 
follows.  Given that the oscillation for $\alpha=0$ is given by
\begin{equation}
\cos\left [\frac{\omega}{\epsilon}\left (\frac{k}{k_*}\right )^\epsilon\right ],
\label{Eqn:argument}
\end{equation}
shifting $\omega$ by $2n\pi\epsilon/(k/k_*)^\epsilon$ for some integer $n$ 
does not change the value of the cosine at wave number $k$.  Furthermore, 
$(k/k_*)^\epsilon$ is a slowly varying function of $\epsilon$, so that 
the value of the cosine is roughly constant for $\omega/\epsilon$ in
integer multiples of $2\pi/(k/k_*)^\epsilon$.  If we now make a simplifying
assumption that the likelihood is dominated by contributions from a 
small range in $k$ around some value $k_d$, then the likelihood function should
exhibit this multiplicity as structures along 
\begin{equation}
\frac{\omega}{\epsilon} = \frac{2n\pi}{(k_d/k_*)^{\epsilon}}.
\label{Eqn:constratio}
\end{equation}

\begin{figure}[ptbh]
\myfigure{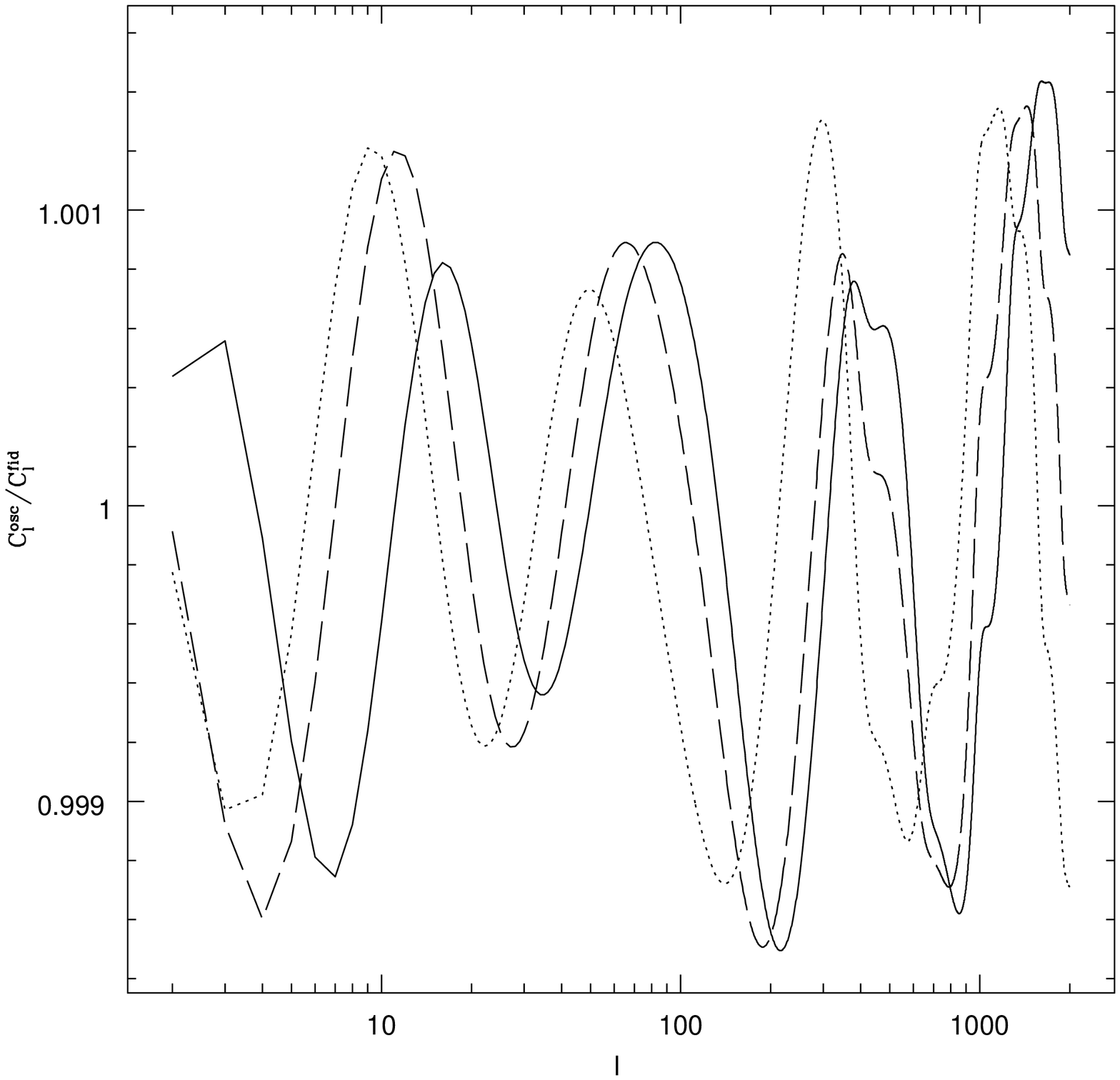}
\caption{Ratio of the oscillatory $C_l$'s to a fiducial $C_l$ without 
oscillations, plotted for three values of $\epsilon$.  Here, $\omega=4.4$,
and the phase is fixed to be zero.  The solid curve denotes $C_l$'s with 
$\epsilon=0.08$, the dotted curve denotes $\epsilon=0.1$, and the dashed curve
denotes $\epsilon=0.12$.  Note the overall coincidence of the extremum 
near $l=200$, where the first acoustic peak lies.}
\label{Fig:threeosc}
\end{figure}

The low likelihood in the range $4<\omega<5$ can be explained by noting the
frequency of oscillatory peaks and troughs in each decade.  In each decade,
there are approximately $\omega\ln(10)/2\pi$ oscillations, so that 
in the range $4<\omega<5$, this corresponds to roughly three 
extrema in the oscillatory features (see Fig.~\ref{Fig:threeosc}).  
The extrema in the 
oscillations tend to coincide with the CMB acoustic peaks, and the overall 
effect of the primordial oscillations is to change the 
heights of the acoustic peaks.  As is well known, the ratio of heights of 
the CMB acoustic peaks can be constrained tightly, and as a result, the 
likelihoods for $4<\omega<5$ tend to be lower.  Because our MCMC sampler moves 
the chain based on the ratio of likelihoods, the lack of high likelihood 
regions between $\omega=4$ and $\omega=5$ implies that the chains do not 
tend to cross into this region, possibly explaining the 
fact that our chains do not explore the region below $\omega=5$.

Given the presence of such irregularities, obtaining reliable constraints on 
the oscillation parameters is difficult.  We believe that our reparametrization 
of the oscillation parameters into physically distinct parameters reduces the 
irregularities in the amplitude parameter, making the results concerning 
$\lambda$ robust.  However, to obtain robust limits on the other parameters,
the presence of multiple maxima needs to be taken into account.

In MCMC analysis, the step sizes are typically taken to be similar to
the characteristic size of the (hopefully global) maximum of the
likelihood.  However, if the local maxima have similar sizes, the
chains can get stuck in the local maxima for a long time and lead to
false conclusions about the location and constraints from the chains.
In exploring these models, it is therefore important to monitor
convergence using multiple chains, since it is unlikely that chains
started from different, randomly chosen starting points would all fall
into the same false local maximum.  Furthermore, nonstandard
algorithms (e.g., high-temperature chains~\cite{Jennison93} and other
modified samplers~\cite{GilksBook}) might be necessary to improve
mixing of the chains, allowing exploration of the irregular features
in a reasonable amount of computation time.

\section{Discussion}
\label{Sect:Discussion}

We have performed a full likelihood analysis of the detectability of 
oscillatory primordial power spectra in current and future CMB data, 
motivated by a phenomenological description of possible effects on the 
primordial power spectrum due to the presence of a high-energy 
cutoff~\cite{ArmendarizLim03}.  Not surprisingly, only a weak constraint can 
be placed on the parameters that define the oscillations, the current data set 
being consistent with a scale-invariant power spectrum without oscillations. 

The analysis of the fundamental limitation from a cosmic variance
limited experiment gives us cause for some cautious optimism,
suggesting that the amplitude of the oscillations can be limited to
less than $0.5\%$.  Under the assumption that the vacuum choice
parameter $|X|\approx 1$, this translates to a possible limit on the
cutoff energy scale of $\Lambda > 200 H_{\text{infl}}$.  Stated differently,
if $\Lambda < 200 H_{\text{infl}}$, the resulting features in the primordial
power spectrum is observable in principle, allowing such a model to be 
distinguished from the null model.  Note that
this is currently the only limit on the cutoff scale using the
effects on the vacuum choice.  As noted above, however, the limits on
the other parameters may not be robust due to the fact that the
likelihood function exhibits multiple local maxima.  Future analyses
may need to take this into account by utilizing MCMC techniques that
facilitate jumps between maxima, such as modifications to the sampling
method.

Is it possible to do better?  At present, the CMB remains the best tool for 
exploring the shape of the primordial power spectrum, limited fundamentally 
by gravitational lensing due to the intervening matter~\cite{HuOkamoto03}.  
Other probes of cosmological fluctuations, such as weak lensing surveys, can
help extend the wave number range that can be probed, but exploring the effect 
of uncertainties from such surveys is beyond the scope of this work.

Given the complex structure of the likelihood function, 
robust results from high-precision surveys will require the use of 
modified MCMC algorithms in order to map out the multiple features in the 
likelihood.  With that caveat, future surveys, such as the Planck and 
CMBPol missions, should be able to limit the shape of the primordial power 
spectrum and provide an insight into the nature of the high energy cutoff.

\begin{acknowledgments}
We acknowledge the use of transfer functions from the hierarchy code
written by W. Hu.  We thank C. Armendariz-Picon, 
S. Carroll, W. Hu, J. Martin, D. Nagai, and J. Ruhl for useful 
comments and discussions. 
T.O. was supported by NASA NAG5-10840 and the US DOE OJI program.
E.A.L. was supported by the US DOE grant DE-FG02-90ER40560 and 
the David and Lucille Packard Foundation. 
\end{acknowledgments}

\newpage
\printtables

\printfigures

\begin{thebibliography}{}
\bibitem{Bennett03} C. L. Bennett \textit{et al.}, Astrophys. J. Suppl. 
\vol{148}, 1 (2003).
\bibitem{Spergel03a} D. N. Spergel \textit{et al.}, Astrophys. J. Suppl. 
\vol{148}, 175 (2003).
\bibitem{Peiris03} H. V. Peiris \textit{et al.}, Astrophys. J. Supp. 
\vol{148}, 213 (2003).
\bibitem{Guth81} A. H. Guth, \prd \vol{23}, 347 (1981).
\bibitem{BST83} J. M. Bardeen, P. J. Steinhardt, and M. S. Turner,
\prd \vol{28}, 679 (1983).
\bibitem{Starobinsky82} A. A. Starobinsky, Phys. Lett. B \vol{117}, 175 (1982).
\bibitem{Planck} \texttt{http://astro.estec.esa.nl/Planck}
\bibitem{CMBPol} \texttt{http://universe.gsfc.nasa.gov/be/roadmap}
\bibitem{MukherjeeWang03} P. Mukherjee and Y. Wang, \apj \vol{593}, 38 (2003).
\bibitem{WangMathews02}  Y. Wang and G. Mathews, \apj \vol{573}, 1 (2002).
\bibitem{Wang99} Y. Wang, D. N. Spergel, and M. A. Strauss, \apj \vol{510},
20 (1999).
\bibitem{BridleLewis03} S. L. Bridle, A. M. Lewis, J. Weller, and 
G. Efstathiou, MNRAS Lett. \vol{342}, 72 (2003).
\bibitem{Miller02} C. J. Miller, R. C. Nichol, C. Genovese, and L. Wasserman,
\apj \vol{565}, 67 (2002).
\bibitem{Tegmark02} M. Tegmark and M. Zaldarriaga, \prd \vol{66}, 103508 
(2002).
\bibitem{HuOkamoto03} W. Hu and T. Okamoto, \prd \vol{69}, 043004 (2004). 
\bibitem{Burgess02} C.~P.~Burgess, J.~M.~Cline, F.~Lemieux and R.~Holman,
JHEP \vol{0302}, 048 (2003).
\bibitem{Brandenberger01A} R. H. Brandenberger and J. Martin, 
Mod. Phys. Lett A \vol{16}, 999 (2001).
\bibitem{Brandenberger01B} J. Martin and R. H. Brandenberger, \prd \vol{63}, 
123501 (2001).
\bibitem{Danielsson02A} U. H. Danielsson, \prd \vol{66}, 023511 (2002). 
\bibitem{Easther02} R. Easther, B. R. Greene, W. H. Kinney  and G. Shiu, 
\prd{66} 023518 (2002).
\bibitem{ArmendarizLim03} C. Armendariz-Picon and  
E. A. Lim, J. Cosmol. Astropart. Phys. \vol{12}, 06 (2003). 
\texttt{hep-th/0303103}.
\bibitem{Elgaroy03}  \O . Elgar\o y and S. Hannestad, 
\prd \vol{68}, 123513 (2003).
\bibitem{Martin03} J. Martin and C. Ringeval, \prd (to be published), 
\texttt{astro-ph/0310382} (2003).
\bibitem{Bergstrom02} L. Bergstr\"{o}m and U. H. Danielsson, JHEP \vol{0212}, 
038 (2002). 
\bibitem{LythLiddleBook} A. R. Liddle and D. H. Lyth, 
\emph{Cosmological Inflation and Large-Scale Structure} (Cambridge University 
Press, Cambridge, U.K., 2000).
\bibitem{Mukhanov91} V. F. Mukhanov, H. A. Feldman, and R. H. Brandenberger, 
Phys. Rep. \vol{215}, 203 (1991).
\bibitem{BirrellDavies} N. D. Birrell and P. C. W. Davies, 
\emph{Quantum Fields in Curved Space} (Cambridge University Press, Cambridge,
U.K., 1982).
\bibitem{Jacobson93} T. Jacobson, \prd \vol{D48}, 728 (1993).
\bibitem{ISW} R. K. Sachs and A. M. Wolfe, \apj \vol{147}, 73 (1967).
\bibitem{Kogut03} A. Kogut, \textit{et al.}, Astrophys. J. Supp.
\vol{148}, 161(2003).
\bibitem{Seljak96} U. Seljak, \apj \vol{463}, 1 (1996).
\bibitem{Zaldarriaga98} M. Zaldarriaga and U. Seljak, \prd \vol{58}, 023003 
(1998).
\bibitem{Hu00}W. Hu, \prd \vol{62}, 043007 (2000).
\bibitem{CMBFAST} U. Seljak and M. Zaldarriaga, \apj \vol{469},
437 (1996).
\bibitem{CAMB} A. M. Lewis, A. Challinor, and A. Lasenby, \apj
\vol{538}, 473 (2000).
\bibitem{LewisBridle02} A.  Lewis and S. Bridle,  \prd \vol{66} 103511 
(2002).
\bibitem{Verde03} L. Verde, \textit{et al.}, Astrophys. J. Supp. 
\vol{148},195 (2003).
\bibitem{GilksBook} W. R. Gilks, S. Richardson, and E. J. Spiegelhalter,
\textit{ed.}, \emph{Markov Chain Monte Carlo in Practice} (Chapman and Hall,
 London, U.K., 1996). 
\bibitem{Metropolis53} N. Metropolis \textit{et al.}, J. Chem. Phys. \vol{21},
1087 (1953).
\bibitem{Hastings70} W. K. Hastings, Biometrika \vol{57}, 97 (1970).
\bibitem{GelmanRubin92} A. Gelman and D. B. Rubin, Statist. Sci. \vol{7}, 457 
(1992).
\bibitem{ACBAR} C.-L. Kuo, \textit{et al.}, BAAS \vol{34}, 1324 (2002).
\bibitem{CBI} T. J. Pearson, \textit{et al.}, \apj \vol{591}, 556(2003).
\bibitem{SZ} R. A. Sunyaev and Y. B. Zel'dovich, Astrophys. Sp. Phys. \vol{4},
173 (1972).
\bibitem{CBIexcess} B. S. Mason, \textit{et al.}, \apj \vol{591},540 (2003).
\bibitem{Jennison93} C. Jennison, J. R. Statist. Soc. B \vol{55}, 54 (1993).
\end{thebibliography}
\end{document}